\documentclass[aps, superscriptaddress, preprint, showpacs, showkeys, preprintnumbers, amsmath, amssymb, longbibliography,nofootinbib ]{revtex4-1}
\usepackage{graphicx}
\usepackage{float}
\usepackage{xcolor}
\usepackage{upgreek}

\begin{document}
\title{The Local Incompressibility of Fractional Quantum Hall States at a Filling Factor of 3/2}

\author{L.~V. Kulik}
\affiliation{Institute of Solid State Physics Russian Academy of Sciences Chernogolovka, Moscow District, 2 Academician Ossipyan Street, 142432 Russia}
\affiliation{National Research University Higher School of Economics, Moscow, 20 Myasnitskaya Street, 101000 Russia}
\author{V.~A. Kuznetsov}
\email{To whom correspondence may be addressed. volod\_kuzn@issp.ac.ru}
\affiliation{National Research University Higher School of Economics, Moscow, 20 Myasnitskaya Street, 101000 Russia}
\affiliation{Institute of Solid State Physics Russian Academy of Sciences Chernogolovka, Moscow District, 2 Academician Ossipyan Street, 142432 Russia}
\author{A.~S. Zhuravlev}
\affiliation{Institute of Solid State Physics Russian Academy of Sciences Chernogolovka, Moscow District, 2 Academician Ossipyan Street, 142432 Russia}
\author{V. Umansky}
\affiliation{Braun Center for Submicron Research, Weizmann Institute of Science, 234 Herzl Street, POB 26, Rehovot 76100, Israel}
\author{I.~V. Kukushkin}
\affiliation{Institute of Solid State Physics Russian Academy of Sciences Chernogolovka, Moscow District, 2 Academician Ossipyan Street, 142432 Russia}
\affiliation{National Research University Higher School of Economics, Moscow, 20 Myasnitskaya Street, 101000 Russia}
	
\begin{abstract}
We studied neutral excitations in a two-dimensional electron system with an orbital momentum $\Delta M = 1$ and spin projection over magnetic field axis $\Delta S_z = 1$ in the vicinity of a filling factor of 3/2. It is shown that the 3/2 state is a singular point in the filling factor dependence of the spin ordering of the two-dimensional electron system. In the vicinity of $\nu=3/2$, a significant increase in the relaxation time ($\tau = 13$ $\upmu\text{s}$) for the excitations to the ground state is exhibited even though the number of vacancies in the lowest energy level is macroscopically large. The decrease of the relaxation rate is related to the spin texture transformation in the ground state induced by spin flips and electron density rearrangement. We claim the 3/2 state is a locally incompressible fractional quantum Hall state.
\end{abstract}

\date{\today}
\maketitle

\section*{Introduction}
The interest in excitations in a two-dimensional electron system (2DES) placed in a quantizing magnetic field is due to their use in a number of fundamental applications. These applications are based on the progress of topological quantum calculations \cite{Witzel2006, Kitaev2003}, the involvement of anyons (quasi particles with neither Bose nor Fermi statistics, and at some filling factors, even being non-abelian) to describe the physical properties of 2DES \cite{Wilczek1982, Wilczek1990}, as well as the discovery of new condensed states that have no obvious analogs in three-dimensional systems \cite{Laughlin1983}. Non-abelian excitations are discussed in relation to fractional quantum Hall states with even denominators. A topological quantum computer, an extremely attractive idea for computation protected from mistakes caused by quantum state decoherence, can be realized using non-abelian anyons \cite{Nayak2008}. The first fractional quantum Hall state of this type, 5/2, was found in experiments on magnetotransport in high-mobility two-dimensional GaAs/AlGaAs-based systems \cite{Willett1987, Willett2009}. More stable at first glance, fractional quantum Hall state at $\nu = 3/2$ was not observed in GaAs until recently \cite{Zhang2009, Fu2019}. Here we do not consider intricate electron systems with complementary degrees of freedom: a layer index in double electron layers or a size-quantized subzone in wide quantum wells. The 3/2 state in GaAs/AlGaAs heterostructures was described as similar to the 1/2 state of composite fermions~\cite{Halperin1983, Halperin1993}, i.e. a partially spin-polarized state of composite fermions with a zero effective magnetic field and a hierarchy of fractional states originating in the 3/2 state as a hierarchy of Integer Quantum Hall states of composite fermions \cite{Du1995, Murthy2003}. Here lies the difference between the 1/2 and 3/2 states, on the one hand, and the 5/2 state, on the other hand, since the 5/2 state is regarded as fully spin-polarized  \cite{Tiemann2012}, which is a necessary condition for the formation of "anti-Pfaffian" and similar type ground states \cite{Moore1991, Lee2007, Son2015}. This physical view was qualitatively confirmed in magnetotransport studies of the fractional quantum Hall effect in novel highly mobile two-dimensional electron systems based on ZnO/MgZnO heterostructures. Significantly higher Zeeman splitting in ZnO makes it possible to engineer a fully spin-polarized 3/2 state. This can be done by varying the angle of the external magnetic field with respect to the perpendicular to the heterostructure surface. As the electron system becomes spin-polarized, the Hall resistance exhibits the formation of a 3/2 plateau \cite{Falson2015}, which makes the 3/2 state potentially interesting for topological quantum calculations.

All these studies deal with the edge state transport of fractional Hall insulators. The existing theoretical models point to a correlation between the transport properties of bulk and edge states in Hall insulators. Therefore, the transport characteristics of a Hall insulator allow for making far-reaching conclusions on the nature of the bulk ground state \cite{Wen1992}. Unlike in Hall insulators, transport measurements give much less information on the nature of the conducted states in fractional Hall conductors. This advance interest in the direct investigation of the bulk properties of fractional quantum Hall states when the fractional state is a conductor. We have developed a number of optical techniques to probe bulk states of 2DES in the quantum Hall regime, namely, resonant reflection (RR), photo-induced resonant reflection (PRR), and photo-induced photoluminescence which can  all be used for the investigation of bulk states in Hall insulators and in Hall conductors \cite{Kulik2016}. The essence of these techniques reduces to the optical formation of nonequilibrium ensembles of excitons with $\Delta M = 1$ and $\Delta S_z = 1$. Such excitations cannot relax directly to the 2DES ground state with photon or phonon emission if the lowest energy spin Landau sublevel is completely occupied, and the spin-orbital interaction is small, as, for example, in the conductance band of GaAs, (this is discussed in detail in \cite{Kulik2015}). Such a relaxation process is impossible, for instance, at integer filling factor $\nu=2$, which reflects the incompressibility of the electron system in the occupied lowest spin Landau sublevel.

As the filling factor is reduced from $\nu = 2$, the 2DES is depolarized and there are no principle restrictions on the relaxation of photoexcited electrons from both spin sublevels of the first Landau level ($M = 1$, $S_z = \pm 1/2$) to the ground state \cite{Aifer1996}. As a result, $\Delta M = 1$, $\Delta S_z = 1$ collective excitations must be unstable. All the more surprising is the observation of a significant increase of in the relaxation time for such excitations in the 3/2 fractional conducting state. The 3/2 state in GaAs/AlGaAs heterostructures appears to be a much more complex physical object than has been considered. Based on our experimental results we claim that the 3/2 state is an ordered spin configuration (spin texture) in which spin flip excitations with a change of orbital momentum are accompanied by the spin texture transformation. This texture transformation forbids the relaxation of the excited electron even though the mean spin polarization through the sample is much less than unity \cite{Tiemann2012}, i.e., there is a macroscopically large number of vacant places in the ground state of the 2DES suitable for the relaxation of the excited electron spin. In other words, we are not dealing with the incompressibility of the whole 2DES as in the case of filling factor 2, but with a local incompressibility: the inability to return the excited electron to the ground state in close proximity to the excited electron itself.

\section*{Results and discussion}

\begin{figure}
	\centering
	\includegraphics[width=0.5\linewidth]{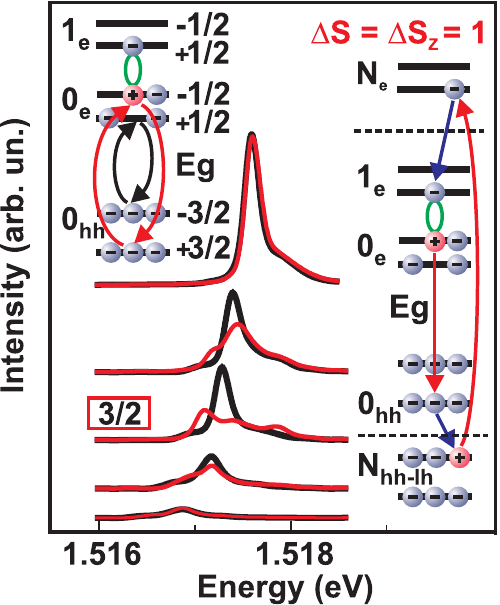}
	\caption{Resonant reflection (RR) (black line) and photoinduced resonant reflection (PRR) spectra (red line) at different filling factors of the 2DES in the vicinity of ($\nu=3/2$) measured at the photoexcitation level $100$ $\upmu\text{W/mm}$. On the right, a diagram of photoexcitation of a nonequilibrium excitation with spin and orbital quantum number 1 is shown including: i) optical transitions from a state in a Landau level of valence band ($N_{hh-lh}$, where hh and lh are heavy and light holes) to a state in a Landau level of the conduction band, ii) hole spin-flip relaxation in the valence band due to the large spin-orbital interaction in the GaAs valence band to the lowest energy state in the zero Landau level of heavy holes ($0_{hh}$), iii) recombination of an electron in the zero Landau level of the conduction band ($0_{e}$) and the hole in the zero Landau level of the heavy hole in the valence band ($0_{hh}$) resulting to a Fermi hole in the upper spin state of the zero Landau level of the conductance band, iv) relaxation of the excited electron with appropriate spin to the lowest spin state of the first Landau level (spin-orbital interaction in the GaAs conductance band is weak), and finally v) binding of the excited electron and the Fermi hole. Examples of the RR and PRR processes are shown on the left side of the diagram.}
	\label{fig1}
\end{figure}

In Fig.~\ref{fig1}, there are examples of the spectra of resonant reflection and photoinduced resonant reflection spectra from the zero Landau level states of the 2DES measured in the quantum Hall conductor in the vicinity of filling factor $\nu=3/2$. At filling factors lower than and above 3/2 (1.3 and 1.7) the RR and PRR signals coincide. This is reasonable as the 2DES is only partially spin-polarized, and there is a macroscopic number of vacant places for the relaxation of the excited electron to the zero Landau level without spin flip and with the emission of a cyclotron energy photon. In this case the 2DES turns into an excited spin state from which it just as easily relaxes to the ground state with the emission of an electron spin resonance photon. These relaxation processes lie in the nanosecond range \cite{Zhang2014, Zhuravlev2014}, so it is not possible to accumulate any considerable number of excitations with such short lifetimes and the photoexcitation density used in the experiment.

\begin{figure}
	\centering
	\includegraphics[width=0.5\linewidth]{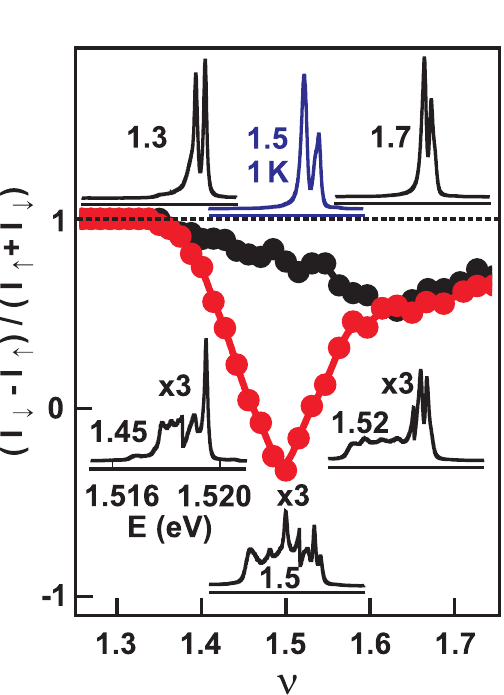}
	\caption{$\alpha=(I\downarrow-I\uparrow)/(I\downarrow+I\uparrow)$ measured from the integral intensity of resonant reflection (RR) (black dots) and photoinduced resonant reflection (PRR) (red dots) lines as a function of filling factor.The insets show photoinduced photoluminescence spectra measured in the same energy interval at the temperature of 0.45~K (black lines) and t different filling factors as well as a spectrum measured at the temperature of 1.5~K at filling factor 3/2. Spectra at $\nu<1.3$ and $\nu>1.6$ as well as the spectrum at 1.5~K consist of two lines corresponding recombination of valence band holes ($0_{hh}$) with the conduction band ($0_{e}$) active in two polarizations of the emitted light. When the nonequilibrium ensemble of excitations with orbital and spin quantum number 1 is created at $\nu>1.3$ and $\nu<1.6$ and a temperature $T<0.8 K$ a variety of new lines appears in the photoluminescence spectra.}
	\label{fig2}
\end{figure}

With a decreasing 2DES filling factor the number of unoccupied places at the zero Landau level increases and the RR and PRR signals grow accordingly. However, of principal interest is not the absolute value of the reflection signal but the distribution of vacant sites between the two spin states in the zero Landau level (g-factor is negative in GaAs): $\kappa=(I\downarrow-I\uparrow)/(I\downarrow+I\uparrow)$ (Fig.~\ref{fig2}). The ratio between the empty places in the upper and lower spin Landau sublevels decreases monotonically from filling factor 1.6 to filling factor 1 according to the RR signal. At filling factor 1, the lower spin Landau sublevel is occupied, and the upper spin Landau sublevel is empty (Hall ferromagnet). Therefore, the observed PRR pattern of the empty place distribution between two spin-sublevels is natural.

\begin{figure}
	\centering
	\includegraphics[width=0.5\linewidth]{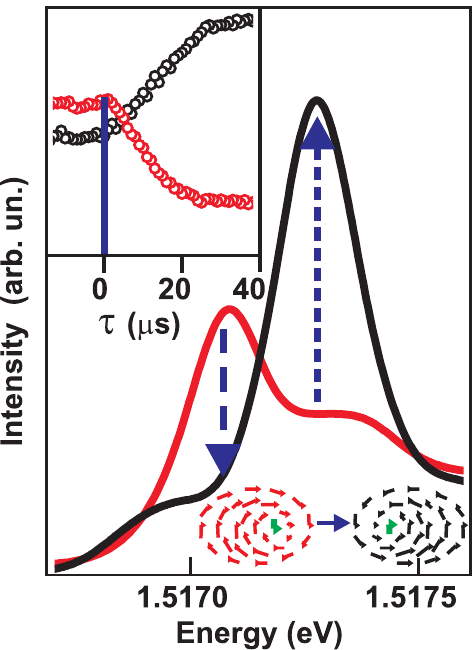}
	\caption{Resonant reflection (RR) (black line) and photoinduced resonant reflection (PRR) spectra (red line) measured at filling factor 3/2. The inset shows the relaxation dynamics of the PRR spectra after photoexcitation switch-off. Below is a schematic illustration of spin texture transformation associated with photoexcitation-induced changes of the spin and orbital quantum numbers of the electron system.}
	\label{fig3}
\end{figure}

Surprisingly, the reflection signal under non-resonant exitation is non-zero in the vicinity of filling factor 3/2 due to the creation of nonequilibrium spin excitations (Fig.~\ref{fig2}). The number of unoccupied states in the lower spin Landau level increases dramatically during the formation of a nonequilibrium spin excitation ensemble. Simultaneously, the lifetime of nonequilibrium spin excitations reaches values over $10$ ${\upmu}\text{s}$ which is only an order of magnitude less than the lifetime of analogous excitations in the Hall insulator at filling factor 2 (Fig.~\ref{fig3}). At filling factor 2, the 2DES is in the incompressible Integer Quantum Hall state with a large energy gap at the Fermi level, while the 3/2 state is a Hall conductor with no energy gap at the Fermi level at all. Even in the incompressible Hall ferromagnet state at $\nu = 1$, the relaxation of excitations with orbital momentum $\Delta M$ and spin projection $S_z = 1$ proceeds much faster than at filling factor 3/2. It occurs despite the number of empty places at the lower spin sublevel in the Hall ferromagnet being much less than in the 3/2 state (yet it is not exactly equal to zero at experimentally achieved temperatures \cite{Plochocka2009}).

It is possible to estimate roughly the nonequilibrium density of excitations with $\Delta M = 1$ and $\Delta S_z = 1$ at filling factor 3/2 as we know of the dependencies of the nonequilibrium excitation density on the photoexcitation density and on the excitation relaxation time obtained from the studies of the Hall insulator $\nu=2$ \cite{Kulik2015, Kulik2016}. At the maximum allowed photoexcitation density causing no overheating of the electron system, 100~$\upmu\text{W/mm}^2$, this value is about 1 \% of the total magnetic flux quanta at a single Landau level. If we assume that excitation effectiveness does not depend on filling factor then excitation density depends only on lifetime. For the filling factor $\nu = 2$, maximum of the excitation density is 0.1 of LL, and lifetime is $100 {~\upmu}\text{s}$. At $\nu = 3/2$ lifetime is $10 {~\upmu}\text{s}$, then excitation density is about 0.01 of the LL.

From Fig.~\ref{fig2} the rough estimation of spin flips in the 2DES ground state is about 27 per excited electron; i.e. the formation of a single spin excitation with $\Delta M = 1$ and $\Delta S_z = 1$ is followed by the rearrangement of the $\sim27$ spins in the ground state. Assuming that PRR intensity $I$ linearly depends on excitation density $n$, we can estimate number of excited electrons in one excitation. Thus, dimensionless degree of spin polarization $\kappa = \frac{I_{\downarrow} - I_{\uparrow}}{I_{\downarrow} + I_{\uparrow}} = \frac{n_{\downarrow} - n_{\uparrow}}{n_{\downarrow} + n_{\uparrow}}$ In the equlibrium, $\kappa = 0.79$. Under photoexcitation, $\kappa = -0.34$. This gives us $\Delta n = 0.28$. Knowing excitation density, we get 27 electrons per one excitation as an estimate. Accordingly, relaxation to the ground state should also accompanied $\sim27$ spin flips and the spin texture rearrangement to the ground state. Despite the metallic conductivity of the 2DES at $\nu=3/2$, we claim that the electrons are united into large spin textures (much larger than those existing in the vicinity of Hall ferromagnet $\nu=1$ \cite{Sondhi1993, Fertig1994}). As a consequence, a physical phenomenon similar to incompressibility of the 2DES emerges in the proximity of the photoexcited electron (local incompressibility). That is the reason for huge experimental relaxation times for the excited electrons.

\begin{figure}
	\centering
	\includegraphics[width=0.5\linewidth]{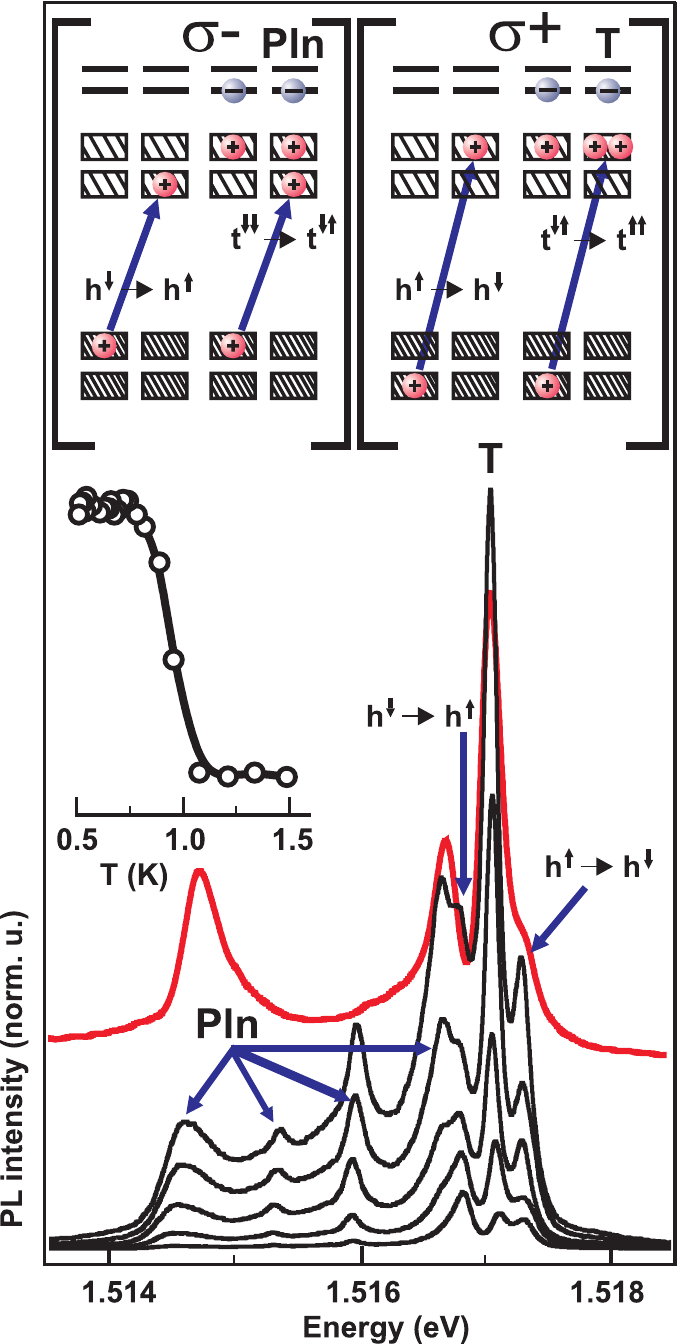}
	\caption{Above shown is a diagram of optical photoluminescence transitions in two polarizations of the emitted photon in the absence and presence of excitation with spin and orbital quantum number 1. Below shown are photoinduced photoluminescence spectra at filling factor 3/2 measured at various photoexcitation levels that are 1, 1.5, 3, 6 and 16 times less than the maximal photoexcitation level $100$ $\upmu\text{W/mm}$ (black solid lines). For comparison, a photoinduced photoluminescence spectrum measured at filling factor 2 at an excitation level of $10$ $\upmu\text{W/mm}$ (red solid line) is presented. The energy scales are equal for black and red lines. The inset shows the ratio of the integral intensity of the plasmaron line to the integral of photoluminescence signal.}
	\label{fig4}
\end{figure}

Not least important criterion of formation of a nonequilibrium excitation with $\Delta M = 1$ and $\Delta S_z = 1$ is the appearance of recombination lines of three-particle complexes composed of a photoexcited hole and nonequilibrium excitation itself in the 2DES photoluminescence spectrum. In the carefully studied case of a Hall insulator at $\nu = 2$, there are two types of such complexes defined by the spin of the photoexcited hole (Fig.~\ref{fig4}) \cite{Zhuravlev2016}. After recombination, the photoexcited hole turns to a vacancy in the filled Landau level of the conduction band (a Fermi hole). Depending on the two Fermi holes spins there appear two bound three-particle complexes built of the nonequilibrium excitation and the Fermi hole, one appearing because of the recombination and another confined in the nonequilibrium excitation. If the spins of two Fermi holes form a spin triplet, the three-particle complex is a trion. If the spins of two Fermi holes form a spin singlet, the three-particle complex is a plasmaron since the excited electron may recombine with one of the Fermi holes transferring energy and momentum to a new electron-hole pair (plasma oscillation). The plasmaron can be regarded as a magnetoplasmon \cite{Bychkov1981} bound to an extra Fermi hole.

The intensity of the trion line accounts for the whole density of the nonequilibrium excitations with $\Delta M = 1$ and $\Delta S_z = 1$ in the excitation spot. Plasmaron photoluminescence contains information on the total number of excitations as well as on the plasmarons  spectral density that yields, in turn, the density distribution of nonequilibrium excitations in the momentum space before they bind to the Fermi hole \cite{Zhuravlev2016, Kuznetsov2018}. The model is only approximately valid for the fractional ground state. Interaction of three-particle complexes with surrounding Fermi holes or spin textures as in the case of $\nu=3/2$ must be taken into account.

The trion and plasmaron lines are observed in the vicinity of filling factor 3/2 after the formation of the nonequilibrium ensembles of excitations with orbital and spin quantum number 1 (Fig.~\ref{fig2}). The trion line at filling factor 3/2 is similar to that at filling factor 2. Yet, the plasmaron line is significantly modified. The plasmaron spectrum exhibits extra maxima that are absent in the spectrum of plasmaron obtained at filling factor 2 (the lines in the plasmaron spectrum at filling factor 2 correspond to the high-density states on the dispersion curve of magnetoplasmons: maxima and minima \cite{Kallin1984}).
An increase in photoexcitation power density changes the ratio between one- and three-particle recombination channels. However, the plasmaron spectrum at $\nu=3/2$ does not display any changes. That means we observe a single particle distribution of magnetoplasmons in the momentum space, and this is radically different from the case of $\nu=2$ \cite{Zhuravlev2016}, where a change in excitation density is accompanied with the redistribution of excitations in the momentum space. With increasing temperature, the photoluminescence of three-particle complexes and the PRR signal disappear threshold-wise simultaneously (Fig.~\ref{fig2},~\ref{fig4}). The characteristic transition temperature is 0.8~K that is much higher than the temperature of composite fermion formation. Yet, it is twice less than the Zeeman activation gap (1.6~K) pointing out on the spin nature of the observed phenomenon.

Our experiments qualitatively support the transport measurements of a spatially confined 2DES at $\nu=3/2$ \cite{Fu2019}. The 2DES incompressibility revealed by Zhang and by Fu and coauthors may occur due to the local incompressibility of the 3/2 fractional state and commensurability of the constriction in the confining potential with the size of one or few spin textures. The confined geometry was proposed by Wen as a powerful tool to probe internal topological order of FQH states~\cite{Wen1992}. One may view the charge current at $\nu=3/2$ being performed not by separate electrons but spin textures. When the constriction size of the Hall conductor becomes commensurable with the spin texture size, the transport in the bulk of the 2DES is terminated, and there remains the edge channel transport only. As a result, the $R_{xy}$ is quantized at $\nu=3/2$ in the constriction. The fractional state 3/2 in spatially confined geometry is, therefore, an example of locally incompressible fractional quantum Hall states being neither Laughlin liquid nor Integer Quantum Hall states of composite fermions, originating from the spin degree of freedom of 2DEG.

\section*{Conclusion}

Optical techniques were used to investigate excited states with orbital momentum $\Delta M = 1$ and $\Delta S_z = 1$ near fractional filling factor 3/2 in a 2DEG. Huge increase in the excitation relaxation time ($\tau = 13$ $\upmu\text{s}$) was discovered. We claim that the spin field is spatially redistributed owing to the existence of an excited electron and a corresponding vacancy. The electrons are united in spin textures with the different spin configurations of the excited and ground states.

\section*{Experimental setup}
\section{Samples}
The experiment was performed using high quality heterostructures with a symmetrically doped single GaAs/AlGaAs quantum well with electron concentration in the two-dimensional channel from $1.8\cdot 10^{11}$ cm$^{-2}$ to $2.6\cdot 10^{11}$ cm$^{-2}$ and dark mobility more than $1.5\div4.0\cdot 10^7$ cm$^2/$V$\cdot s$. Symmetric doping was required to minimize the penetration of the electron wave function into the barrier of the quantum well. The quantum well widths were $30\div40$~nm.
Since there is qualitative agreement between the experimental results for different heterostructures, we focus here on the results obtained for heterostructures with a quantum well width of 31~nm and a concentration of $1.8\div2.1\cdot 10^{11}$ cm$^{-2}$.

\section{Experimental setup}
The samples in question were placed into a pumped cryostat with liquid $^3$He which in turn was put into a cryostat with a superconducting solenoid. Optical measurements were made in the temperature range of 0.45--1.5~K.

In the experiments we used a double optical fiber technique. One fiber served for the continuous wave resonant and non-resonant photoexcitation of the 2DES. The size of the photoexcited spot was 1~mm$^2$.
The other fiber was used to collect the resonant reflection signal and the photoluminescence signal from the photoexcitation spot, as well as for transferring the signals to the entrance slit of a grating spectrometer equipped with a CCD camera. A broadband laser diode (excitation wavelength 780 nm and spectral width 10 nm) was used as an optical source to form a nonequilibrium spin excitation ensemble and a photoluminescence signal, where the resonant reflection was obtained with a TOptica tunable diode laser with a spectral bandwidth of 20~kHz.
The relaxation time was measured in the time range of $1\div1000$ ${\upmu}\text{s}$ by modulating the laser diode pump with a pulse generator. The photoexcitation density varied within $1\div100$ $\upmu\text{W/mm}$, which did not cause the overheating of the 2DES. PRR per se does not involve the formation of excitations, yet, the 2DES resonant reflection and photoluminescence signals overlap spectrally.
For measurements of the photoluminescence spectra, the resonant excitation was switched off. The parasitic reflection from the heterostructure surfaces was filtered by crossed wire-grid linear polarizers placed between the optical fibers and the sample.
Since only circularly polarized light from the 2DES is generated in a magnetic field, the resonant reflection and photoluminescence pass through the linear polarizer of the collecting optical fiber.

\section{Data availability}

Data are available from the corresponding author upon request.

\begin{acknowledgments}
The work was supported by Russian Science Foundation, grant \#18-12-00246.
\end{acknowledgments}

\section*{Author contributions}
L.V.K., V.A.K., A.S.Z., and I.V.K. designed experiments. V.U. manufactured samples. V.A.K. and A.S.Z. performed experiments. A.S.Z. processed raw data, L.V.K. prepared figures. L.V.K. and V.A.K. wrote the manuscript. All authors participated in the results discussion.
\section*{Additional information}
\textbf{Competing financial interests:} The authors declare no competing financial interests.

%

\end{document}